# Mechanically induced interaction between diamond and transition metals

Zhijie Wang[1], Susheng Tan[2], M. Ravi Shankar[1]

[1] *Department of Industrial Engineering, Swanson School of Engineering, University of Pittsburgh, PA 15261, United States of America*

[2] *Petersen Institute of Nanoscience and Engineering, University of Pittsburgh, Pittsburgh, Pennsylvania 15260, USA*

**ABSTRACT**

Purely mechanically induced mass transport between diamond and transition metals are investigated using transition thin metal film deposited AFM tip scratching and *in situ* TEM scratching test. Due to the weak strength of the transition metal-diamond joints and transition metal thin films, AFM scratching rarely activated the mass transport interaction at the diamond – transition metal thin film interfaces. *In situ* TEM scratching tests were performed by using a Nanofactory STM holder. The interaction at diamond and tungsten interface was successfully activated by nanoscale in-situ scratching under room temperature. The lattice structure of diamond and tungsten were characterized by HRTEM. The stress to activate the interaction was estimated by measuring the interplanar spacing change of tungsten nanotips before scratching and at the frame that the interaction was activated.

# 1. Introduction

Diamond is an allotrope of carbon where each carbon atom is sp$^3$ hybridized and combined with four adjacent atoms by covalent bonds. This specific structure and the short C-C bond length donors diamond the highest hardness in nature[1], ultra-corrosion resistance[2, 3], excellent thermal stability[4, 5], and other unique properties. Therefore, diamond has been widely used in many industrial fields, such as microelectromechanical systems (MEMS)[6, 7, 8, 9], optical elements[10, 11, 12], sensors[13, 14, 15, 16], power devices[17, 18, 19], and information processing[20]. Diamond surface manufacturing is one of the most essential processes to realize these applications. However, manufacturing diamond surfaces, which is essential for the utilizing of diamond, is extremely difficult via ordinary machine methods due to the extremely high hardness of diamond.

The key focus in existing diamond etching methods has been to accelerate material removal and to enable scalability over large areas within a serial production framework[21, 22, 23, 24, 25, 26]. These methods are reliant on fluidic etchants (i.e., plasma, gas, and liquid solution), which place limitations on the geometric complexity that can be endowed. For example, creating a hierarchical structure will require a multistep mask fabrication to fabricate structures in a sequential fashion. More importantly, these etchants require specialized equipment and handling due to their high reactivity, which magnifies the capital costs and limits their scalability for large-volume production. Previous research and industrial production indicated that diamond tools suffered real fast degradation when cutting transition metals like titanium (Ti), Niobium (Nb) and zirconium (Zr). This phenomenon was considered caused by the interaction between diamond and transition metals triggered by the temperature increasing during the machining process, and the oxygen in atmosphere were also believed to be an important element in the reaction. Previous research in studying diamond turning process in a SEM chamber (residual pressure $< 5 \times 10^{-6}$ torr) at small length scales revealed that rapid degradation of diamond tool cutting edges when cutting transition metals at very low speed (~150μm/s, no significant temperature increasing) in a SEM chamber (high vacuum), and an intimately bonded "built up edge" on the single-crystal diamond tool surface (in contact with deformed chip) was formed and the cutting edge of the diamond tool degraded rapidly when cutting transition metals even at really low speed[27], which indicates the high temperature and oxygen are not necessary for activating the reaction between diamond and transition metals. According to the existence of mass transport between diamond and some transition metals, it's possible to manufacture diamond surface via solid state transition metal etchants. This technique can be thermomechanically driven and provides a designable nanoscale high precision and efficiency 3D diamond surface manufacturing.

The advantage of diffusional mass-transport based approaches for patterning is the ability to eschew damage and implantation from the etching sources (e.g., plasma or ion beam). In addition, the solid-state etchant is localizable to enable one-step hierarchical structuring. Achieving the ability to pattern and localize the material removal can enable μm and sub-μm scale selectivity in material removal, which is necessary for the creation of optical components from diamond (e.g., gratings and waveplates), especially polycrystalline diamond, which is easier and more economical to be grown into a large dimension (>1 in.) for applications that exploit its functional properties. In such applications, scalable utilization of spatially patterned transition metal on diamond surfaces with the aim of replicating ultrafine features becomes feasible. With the availability of an array of compositionally modulated etchants, it is possible to envision site specific patterning of compositionally graded films on diamond, which can replicate hierarchical surface features.

There is few research on diamond surface structuring via solid state diffusional mass-transport based approaches. Nagai et al. realize an anisotropic etching of diamond by thermal-chemical reaction between diamond and Ni in high temperature water vapor. Pyramid structure was formed on a (100) diamond surface after etching due to the etching rate anisotropy. The mechanism of this etching process is believed that the Ni were oxidized into NiO by water vapor, which transferred carbon to $CO/CO_2$ to realize the removal of carbon[28]. Besides water vapor and oxygen (oxidizing gases), hydrogen has also been used as the annealing atmosphere for diamond etching and was considered realizing diamond etching by transferring carbon to $CH_4$, which was catalyzed by transition metals[29, 30]. However, the current research on diamond-transition metal interaction and its application on diamond were insufficient. First, the etching spatial selectivity is rarely controllable. For example, the etching roughness is high (Ra > 1 μm), and the etching rate is strongly reliant on diamond surface grain orientations[28, 31]. Secondly, the mechanism of the diffusional interaction between diamond and transition metals, which is essential to optimize this etching approach, is not totally understood. Last but not the least, the interaction was only thermally induced. Motivated by the ability of d-shell rich transition metal to rapidly react with diamond even in SEM chamber (high vacuum) with an ultra-low cutting speed and cutting depth, the research of mechanically induced only reaction between diamond and transition metal and its application on diamond etching should be investigated. This will be helpful to achieve directly write and more precisely (nanoscale) diamond surface structuring approaches.

In this study, purely mechanically induced mass transport between diamond and transition metals are investigated using transition thin metal film deposited AFM tip scratching and *in situ* TEM scratching test. The interaction at diamond and tungsten interface was successfully activated by nanoscale in-situ scratching under room temperature. The lattice stress for activating the interaction between diamond and metals was estimated by measuring the interplanar spacing change of tungsten nanotips before scratching and at the frame that the interaction was activated at the atomic scale.

## 2. Methods

2.1 Mechanical induced reaction tests using atomic force microscope (AFM)

Atomic force microscope (AFM) is a kind of scanning probe microscope. In a typical AFM, a well-fabricated tip (typically made of Si or $Si_3N_4$) scans over the sample surface, and measures the morphology of the surface by the deflection of the tip cantilever, which caused by the force (Van der Waals force, capillary force, electrostatic force, and repulsive force) changing as a function of tip-sample surface distance[32, 33, 34, 35, 36]. Thanks to the ultra-fine manufactured tip (top diameter < 50nm) and the precision record of the contact force, AFM is able to provide a nano-scale direct writing on diamond surface with in-situ control[37, 38]. During this process, the tip is put in contact with diamond with calculated normal and lateral forces. Mechanical induced reaction tests aim to investigate the effect of the parameters of mechanical loading (normal stress, shear stress, and scratching velocity) on the diffusional mass transition rate between diamond and different transition metals. These tests were performed using an atomic force microscope (AFM, Veeco multimode V in Figure 1b) or in-situ TEM scratching system. Figure 1a shows a sketch of the set up for the tests driven by the AFM. The AFM tips (n-type Si, k= 40N/m) were deposited with transition metal films in the e-beam evaporation system (Ni, FeCoB) or the sputtering system (Nb). The transition metal film coated AFM tips were then driven to scan back and force on the diamond surface on scratching mode. The force can be controlled by the z-distance (the z-axis value between the tip touch diamond surface (no pressure) and the scratching was performed). The diamond surface morphologies and nanostructure after scratching were characterized via AFM and SEM.

2.2 *In situ* TEM mechanical test

The experiment was performed in a Thermo Fisher Titan Themis G2 200 probe Cs corrected scanning transmission electron microscope (STEM) chamber using a Nanofactory scanning tunneling microscope (STM) holder to do nanoscale in-situ diamond - tungsten frictional tests. The holder has one fixed end and one triaxially movable end controlled by a manipulator. Niobium (Nb) and tungsten (W) rods were fabricated into nanotips with flat heads using in-situ welding method[39, 40, 41, 42, 43, 44, 45]. The process was illustrated in Figure 2. First, pure transition metal wires (0.25mm in diameter, 99.99% purity, Thermo Scientific™) were clipped into rods with a length of 3mm. The fracture surfaces have sawtooth structure with nanotips on the top. However, the tips do not have flat head surface for frictional tests (shown in Figure 3a), therefore, in situ welding processes were performed to create flat heads for the tips. As illustrated in Figure 2, a transition metal rod with nanotips were connected with an electric pole while another rod made by the same transition metals were connected with the opposite pole. When the nanotip met with the opposite transition metal rod, it welded with the opposite transition metal rod forming a neck shape joint between the two rods due to Joule heating (Figure 3b). Subsequently, the joint was quenched and recrystallized due to the fast heat dissipation through the substrate[46]. After that, tensile force was applied to the welded rods by the holder, and the monocrystalline transition metal joint fractured, forming a nanotip with flat head (Figure 3c), which was eligible for further friction experiment.

## 3. Results and Discussions

3.1 Mechanical scratch induced by atomic force microscope

Figure 4 shows the SEM images of the AFM tips (Mikro Masch HQ NSC 15, k = 40 N/m) before and after depositing 20 nm thickness of FeCoB thin film. As can be seen, the silicon AFM tip was in a pyramid shape with very sharp top (with a diameter < 50 nm), and the tip shape did not significantly change after FeCoB thin film deposition. The AFM tip was driven by the path-designed mode and direct write the designed path on the surface. Figure 5 shows the AFM top view images of the back side surface of a CD and a diamond surface before and after being scratched by FeCoB coated AFM tips. The AFM tip created a permanent scratch with a perfect replica of the designed path. This result proved that the AFM could realize the goal of direct writing with nanoscale resolution. Subsequently, the FeCoB thin film coated AFM was scratched on diamond surface. The z-axis deflection of the tip is 10 nm and the tip velocity is 100 nm/s. The normal force can be simplified calculated by $F = kx_z = 40\,N/m \times 20\,nm = 800nN$. Based on the SEM images, the tip – surface contact area S is not larger than 50 nm 8× 50 nm = 2500 nm². The normal stress is $\sigma = \frac{F}{S} \geq 320 MPa$. However, the diamond topographies did not show significant change before and after scratching. This indicates that no significant interaction between diamond and FeCoB occurs under this circumstance.

Moving forward, AFM tips were scratched back and forth multiple times with low speed to increase the duration of the diamond-FeCoB interactions. An FeCoB coated AFM tip was scratched diamond surface back and forth for 10 times with a velocity of 20 nm/s. Figure 6a shows the designed path of AFM tip on a diamond surface. To easily locate the scratching area, the scratch was processed beside a natural defect of diamond surface. The defect was shown in the left bottom corner of the AFM top view images of the diamond surface. Figure 6b shows the surface topography after scratching. As can be seen, some residue particles are shown at the endpoints of the tip path. The particle has a diameter of approximate 100 nm and a height of 9 nm. However, there was very little residue shown in the middle of the tip path, which indicates that the residue at the endpoints could be FeCoB stripped from the tip but did not interact with diamond. To further confirm this hypothesis, the sample were further solvothermally cleaned by acid mixture ($HNO_3$ (68 wt%): $H_2SO_4$ (98 wt%) = 3:1). As can be seen, the residues shown in the Figure 6b were totally removed by acid, and the scratched area was as smooth as the unscratched area. This indicates that the interaction between diamond and FeCoB thin film was not activated. Based on the surface profile before acid cleaning (after scratching), the low strength of diamond-FeCoB joint (formed by e-beam evaporation) and the low strength of FeCoB thin films might limit the energy to activate the interaction between diamond and FeCoB. Nb and Ni coated AFM tips were also tested in the same setup, however, there is still lack of evidence of interaction between diamond and transition metal through the AFM scratching methods. To further investigate purely mechanically induced diamond-transition metal interaction, in-situ TEM diamond-transition metal milling tests were performed.

3.2 Investigation via *in situ* TEM Milling

Due to the ultra-rapid degradation of diamond tools when use it to cut Nb under vacuum with a low depth of cut and cut speed, the attempt of fabrication the Nb nanotips with flat heads was performed first. However, the structure of Nb nanotips turns to be amorphous after the in-situ welding process, as can be seen in Figure 7. The amorphous structure of the fabricated nanotip is not desirable for further in-situ friction experiment under HRTEM observation, therefore, W was used as the raw material for fabricating nanotips with flat heads. As can be seen in Figure 3, the W nanotip kept its single crystalline structure after *in situ* welding process.

After the in-situ welding process, the opposite W rods were replaced by diamond nanosheets for in-situ friction tests. The nanosheets used in the research are the edge of commercial CVD diamond powder (synthetic, <1 micron, 99.9% (metals basis), Thermo Scientific™). The microstructure of the diamond powder and their thin edges (nanosheets) are shown in Figure 8. The diamond powder was stuck at one end of a W rod (0.25mm in diameter, 3mm in length) by conductive silver paint (PELCO®) to enable its installment on the Nanofactory scanning tunneling microscope (STM) holder. The assembled holder was subsequently placed into the STEM (Thermo Fisher Titan Themis G2 200 probe Cs corrected) chamber for in-situ frictional test. The diamond was fixed while the W nanotips were movable with a sliding speed of 0.5 nm/s.

Figure 9 shows the diamond crystalline structure before and after scratching with a W nanotip. As can be seen, W nanotip has a body-centered cubic (BCC) single crystalline structure and diamond shows a single crystalline structure. The sliding direction is among [110] direction of W lattice and [001] direction of diamond lattice. Before scratching, the diamond edge shows a single crystalline lattice, as proved by the diffraction pattern in this area (Figure 9b). After scratching with W, the diamond crystalline diffraction pattern disappeared (Figure 9d)) in the corresponding area of Figure 9b, and the structure of the scratched area turned to be amorphous. This indicated that the interaction between diamond and W was initiated by the scratching and proved that the interaction between diamond and transition metal can be purely mechanically activated. The W was compressed along [110] direction during the scratching process, in order to investigate the stress required to activate this interaction, the lattice parameter of W was measured before scraping with diamond and at the frame that the interaction between diamond and W was initiated. The result was shown in Figure 10. As can be seen, the interplanar spacing of the (110) plane and ($\bar{1}$10) plane of the bcc W is 2.240 Å and 2.236 Å, which is quite close to the theoretical interplanar spacing value of the {110} planes of bcc W. The theoretical interplanar spacing value of {110} planes of bcc W is calculated by:

$$d = \frac{a}{\sqrt{h^2 + k^2 + l^2}} = \frac{3.165}{\sqrt{1^2 + 1^2 + 0^2}} = 2.238 \text{Å} \quad (1)$$

Figure 10b shows the W lattice structure close to the scratching interface at the frame that the interaction between diamond and W was initiated. As can be seen, the interplanar spacing of original (110) plane (the orientation of (110) plane slightly changed due to the deformation) of W decreased to 2.202 Å at this frame, where that of original ($\bar{1}$10) slightly decreased to 2.232 Å at this frame. In addition, the angle between the original (110) and ($\bar{1}$10) has slightly increase from 90° to 91°. Assume the strain is within the (001) plane. The strain along x axis: $\varepsilon_x = -1.7\%$, the stain along y axis: $\varepsilon_y = -0.18\%$, and the shearing stain: $\gamma_{xy} = -0.0175 \text{ rad}$.

The value of maximum normal strain can be calculated by[39, 42, 47, 48]:

$$|\varepsilon_{max}| = \left| \frac{\varepsilon_x + \varepsilon_y}{2} - \sqrt{\left(\frac{\varepsilon_x - \varepsilon_y}{2}\right)^2 + \left(\frac{\gamma_{xy}}{2}\right)^2} \right| = 0.021 \quad (2)$$

, and the orientation of maximum normal strain is $\alpha = \frac{tan^{-1}(\frac{\gamma_{xy}}{\varepsilon_x - \varepsilon_y})}{2} = 25°$, which is close to [250] of the W lattice.

The value of maximum shear stain can be calculated by[48]:

$$|\gamma_{max}| = \sqrt{(\varepsilon_x - \varepsilon_y)^2 + \gamma_{xy}^2} = 0.023 \quad (3)$$

, and the orientation of maximum shear strain is $\beta = \frac{tan^{-1}(\frac{\varepsilon_y - \varepsilon_x}{\gamma_{xy}})}{2} = -20°$, which is close to [210] of the W lattice.

The maximum of the normal stress can be calculated by:

$$\sigma_{max} = E_\alpha |\varepsilon_{max}| \tag{4}$$

where $E_\alpha$ is the Young's modulus of W along the direction of $\varepsilon_{max}$ of the W lattice (70°). The Young's modulus can be calculated by[42, 49]:

$$E_{ijk} = \frac{1}{S_{11} - 2\left(S_{11} - S_{12} - \frac{S_{44}}{2}\right)\left(l_{i1}^2 l_{j2}^2 + l_{j2}^2 l_{k3}^2 + l_{i1}^2 l_{k3}^2\right)} \tag{5}$$

where i, j, k is the direction index, $S_{11}$, $S_{12}$, $S_{44}$ are the elastic compliances and are equal to 0.257×10⁻² GPa⁻¹, -0.073×10⁻² GPa⁻¹, and 0.66×10⁻² GPa⁻¹, respectively at room temperature. $l_{i1}$, $l_{j2}$, $l_{k3}$ are the direction cosine values along [ijk], which are equals to 0.342, 0.940, and 0 for [250][49]. The Young's modulus of W along the direction of $\varepsilon_{max}$ can be calculated[39, 42]:

$$E_\alpha = 389 \; Gpa \tag{6}$$

The stress is[42]:

$$\sigma_{max} = E_\alpha |\varepsilon_{max}| = 8.169 \; GPa \tag{7}$$

,which is within the idea elastic strength of W[50]. The direction of $\sigma_{max}$ is close to [250] direction of the W lattice.

The maximum of the shear stress can be calculated by:

$$\tau_{max} = G_\beta |\gamma_{max}| = 3.588 GPa \tag{8}$$

, where $G_\beta = 156 GPa$ at room temperature[51]. The direction of $\tau_{max}$ is close to [210] direction of the W lattice.

Therefore, about 8.169 GPa maximum normal stress with a direction close to [250] and 3.588 GPa maximum shear stress with a direction close to [210] is required to activate the reaction between diamond (with surface orientation in [110]) and W (sliding direction in [110]).

## 4. Conclusion

In this study, two approaches were used to investigate purely mechanically induced diamond and transition metal interaction were performed. First, AFM tips deposited with transition metal thin films were used to scratch diamond surface using contact mode. FeCoB, Ni, and Nb thin films were deposited with a thickness of 20 nm on the AFM tips. Based on the size of the tip head contacted with diamond, the spring constant of the AFM cantilever, and the deflection of the AFM cantilever, the calculated normal force during scratching was about over 320 MPa. However, after scratching, transition metal was stacked at the end of the scratching routes. After solvothermally cleaning the scratched diamond samples by acid mixture ($HNO_3$ (68 wt%): $H_2SO_4$ (98 wt%) = 3:1), the transition metal was totally removed. In addition, the AFM top view images showed that the topographies of the scratched area had not changed, which indicated that no significant interaction between diamond and transition metal was activated under this condition. This might be caused by the limitation of real stress applied by the AFM tips due to the weak strength of the diamond and transition metal joints formed by e-beam evaporation and the weak strength of the transition metal thin film.

Moving forward, an in-situ TEM observation of diamond/W friction experiment was performed to investigate purely mechanically induced diamond and transition metal interaction. The experiment were performed using a Nanofactory STM holder. The W nanotips with a flat head were fabricated by an in-situ welding method, and the diamond nanosheets are the edge of commercial diamond powders (diameter < 1µm). In the experiment, the sliding direction is along [110] direction of W lattice and [001] direction of diamond lattice. After friction, the scratched area of diamond nanosheet changed from single crystalline structure to amorphous lattice structure.

By measuring the interplanar spacing change of W nanotips before scratching and at the frame that the interaction was activated, the stress to activate the interaction was estimated.

**Author Contributions**

Z.W. conducted the *in situ* TEM experiments and performed the experimental data analysis. Z.W., S.T. and M.S. prepared the paper with the contribution of all authors.

**Conflict of interest**

The authors declare no competing financial interest.

Figure Captions

(a) 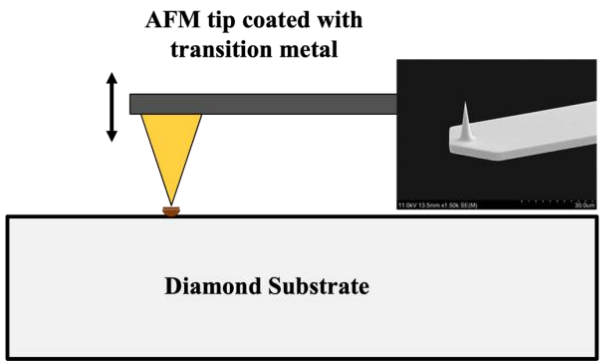 (b) 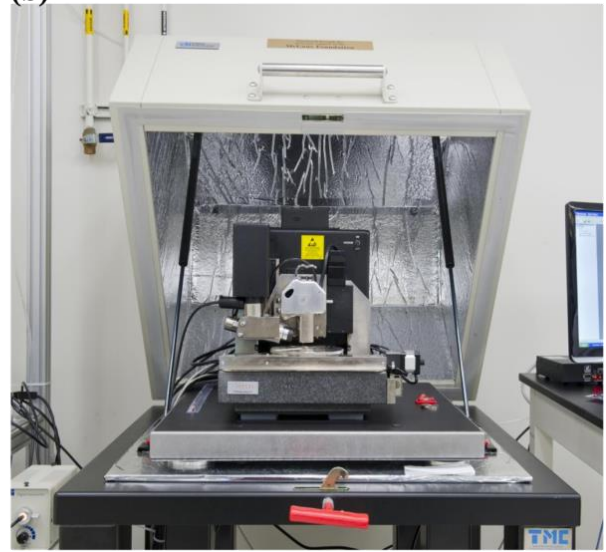

Figure 1. Schematic and equipments of mechanical-induced diamond- transition metal diffusional reaction experiments. (a) Sketch of the tests driven by AFM; (b) Veeco multimode V Scanning Probe Microscope (SPM/AFM).

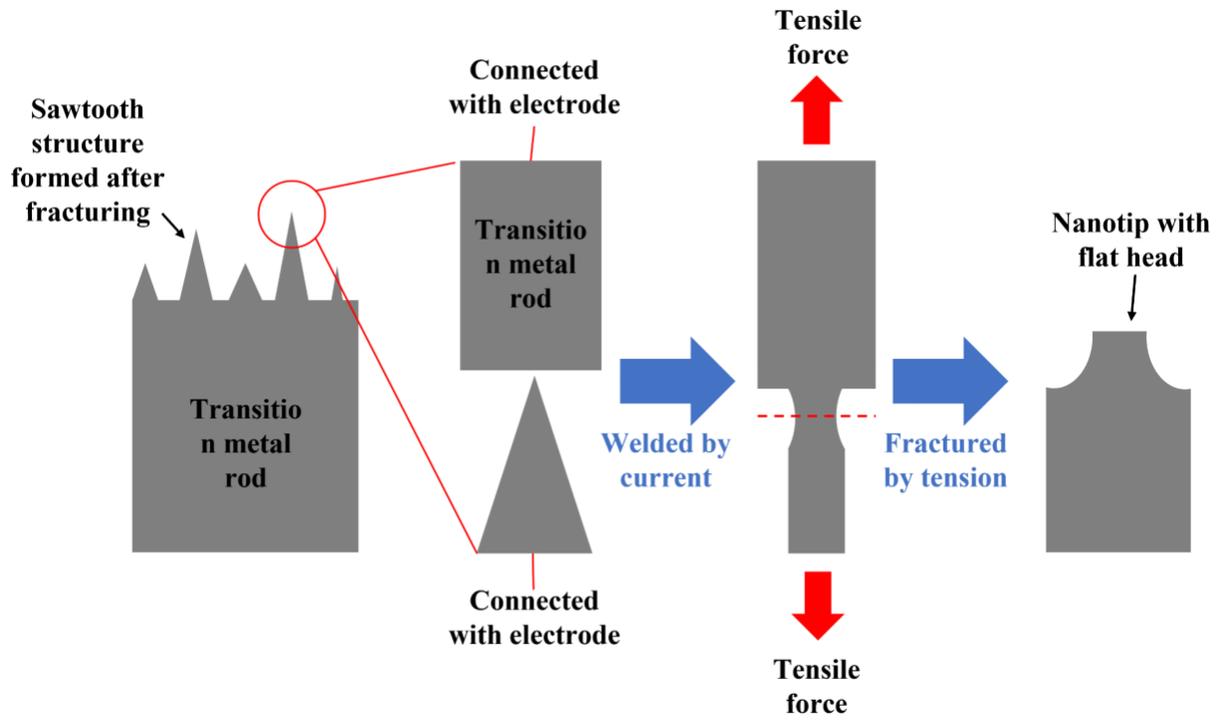

Figure 2. Schematic of in-situ welding process to fabricate transition metal nanotips with flat heads.

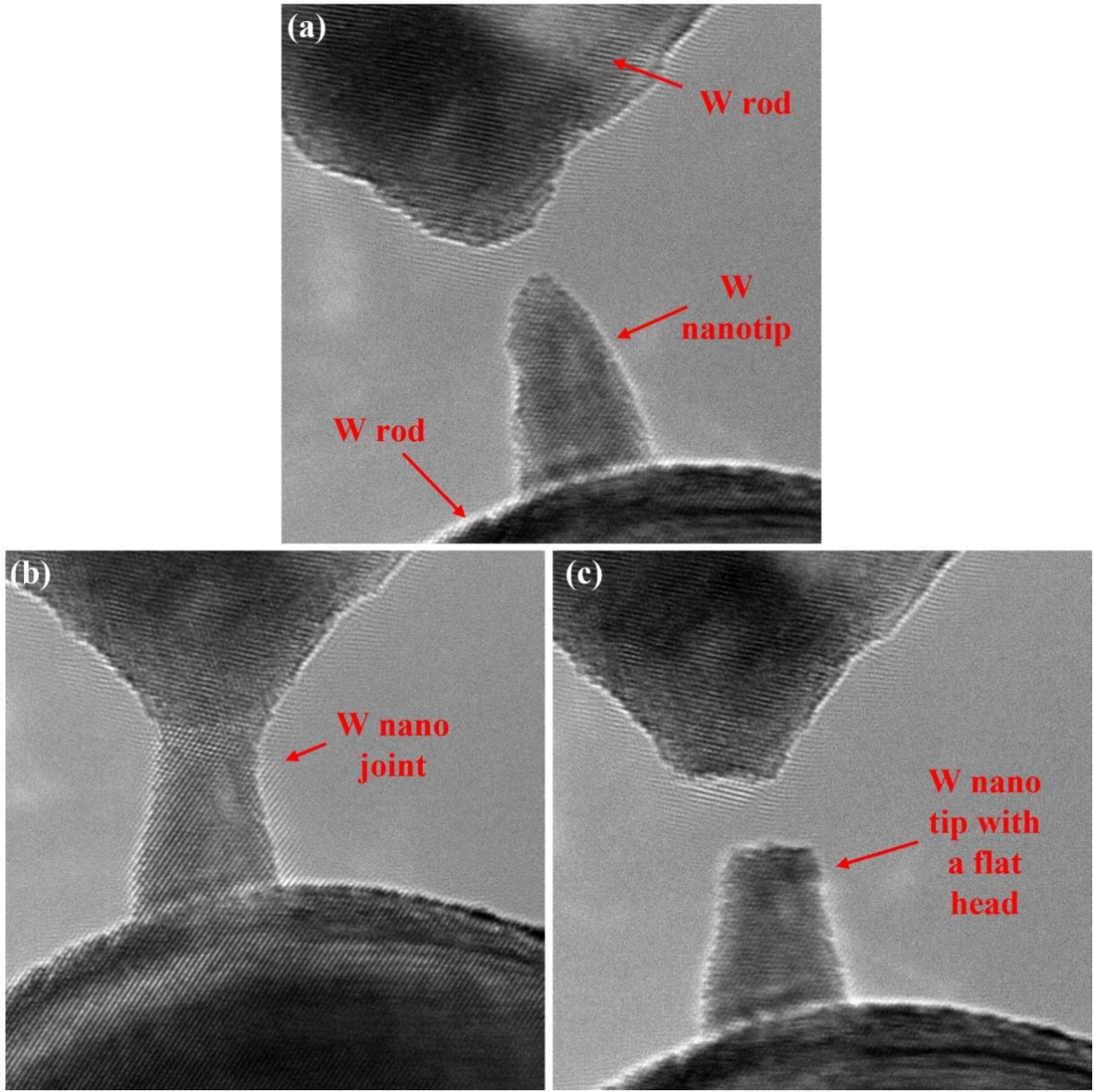

Figure 3. TEM images of in-situ welding tungsten (W) nano tip fabrication. (a) before welding, (b) a W nano joint formed due to Jouel heating, (c) a W nano tip with a flat head created after tensile fracture.

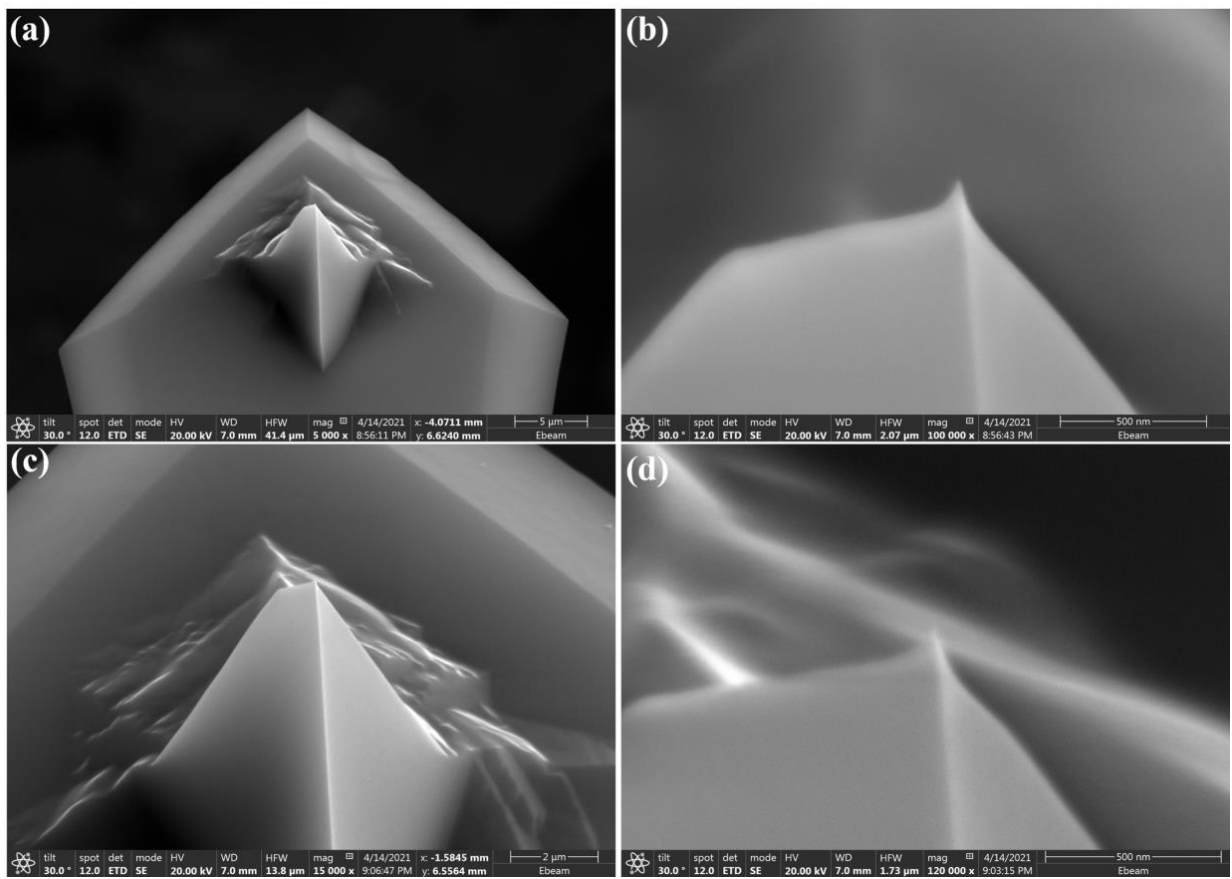

Figure 4 SEM top and side view images of the AFM tips (a) and (b) before and (c) and (d) after deposition of 20 nm FeCoB.

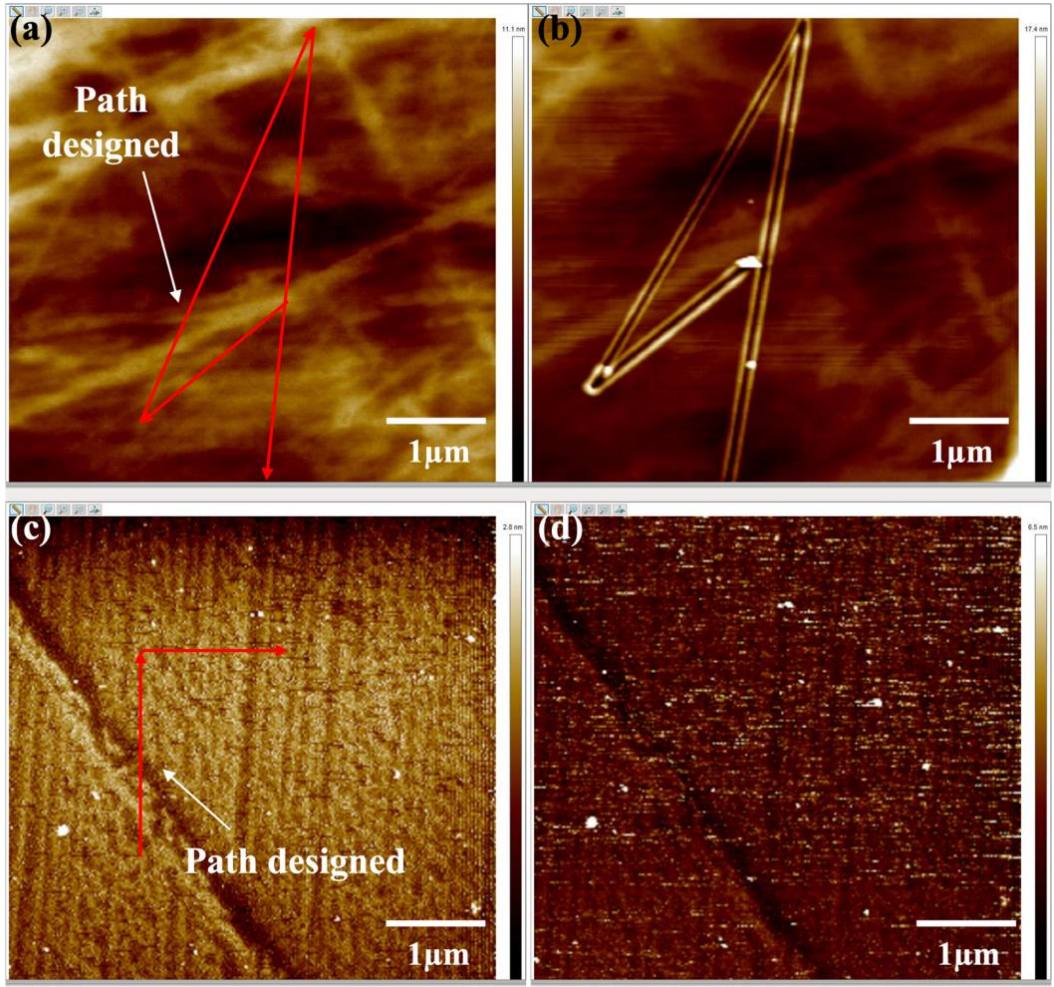

Figure 5 AFM top view images of the back side surface of a CD (a) before and (b) after scratching, diamond surface (c) before and (d) after scratching

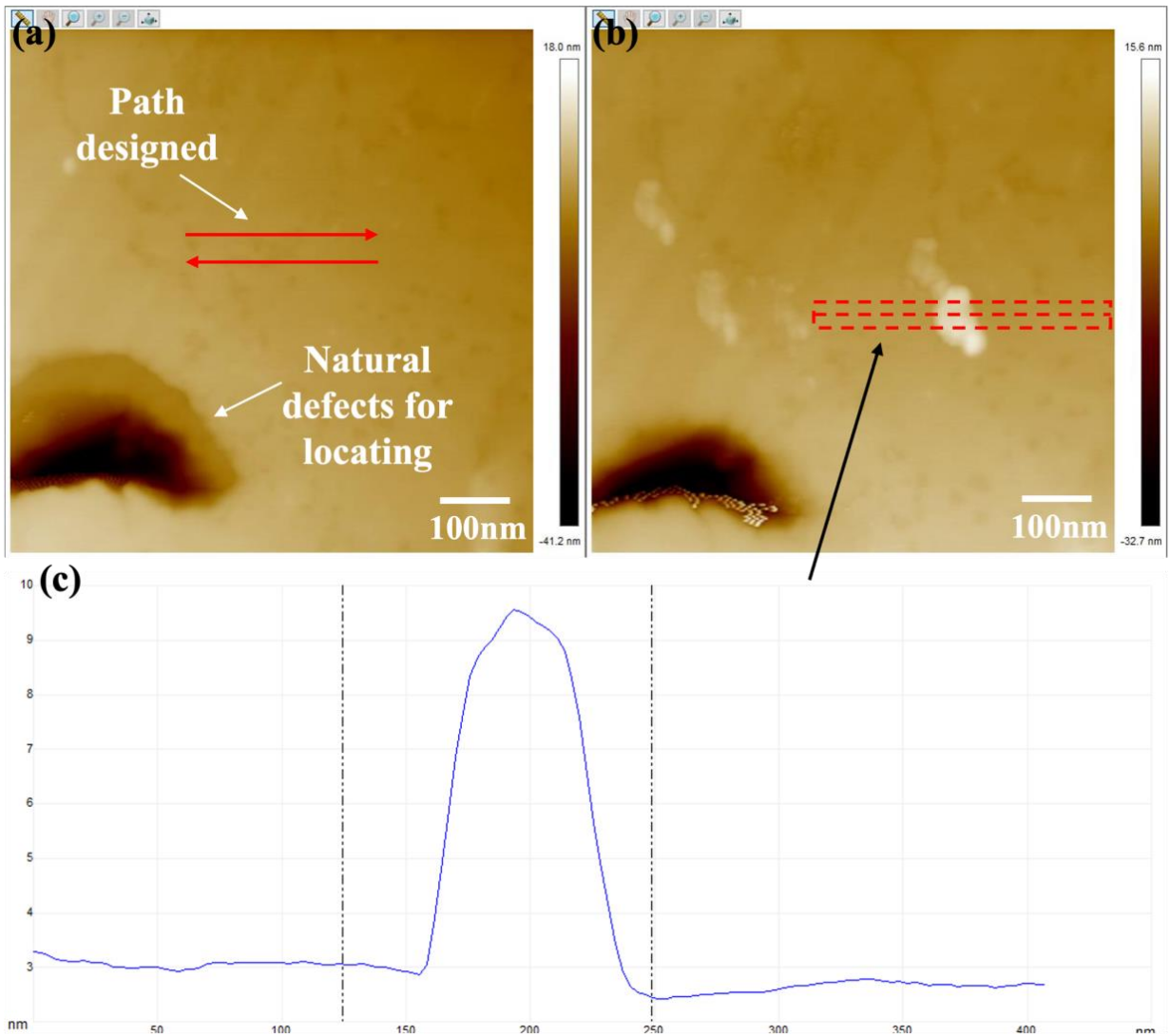

Figure 6 AFM top view images of diamond surface (a) before and (b) after scratched by an FeCoB coated AFM tip. (c) Height profile of the scratching residue shown in (b).

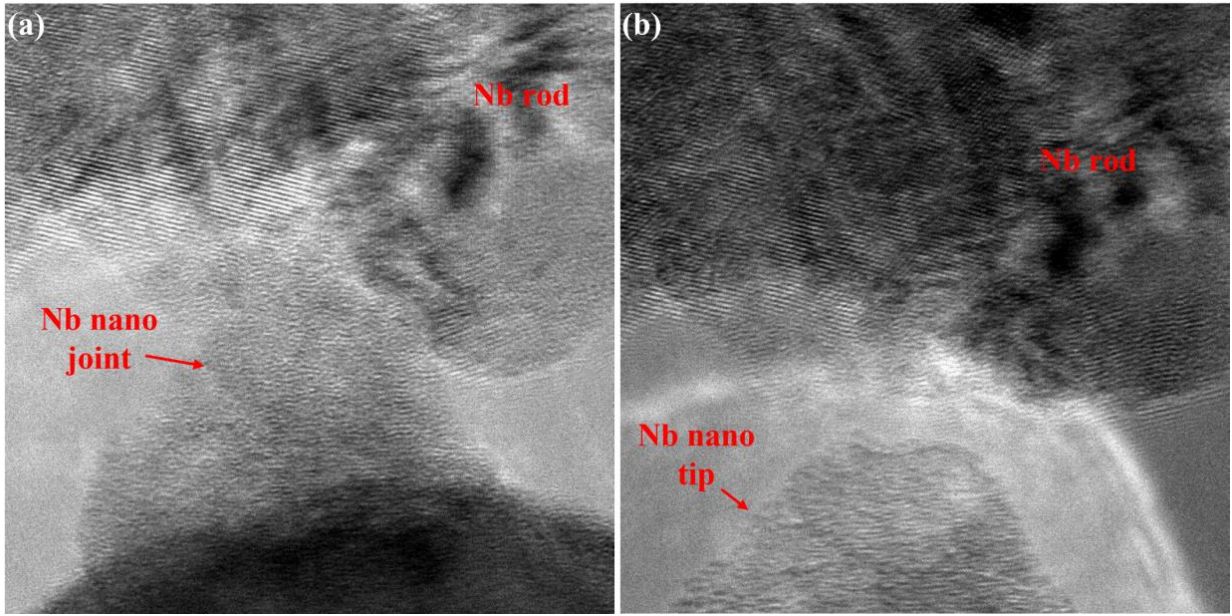

Figure 7. HRTEM image of in-situ welding niobium (Nb) nano tip fabrication. (a) a Nb nano joint formed due to Jouel heating. (b) Nb nanotip after tensile fracture.

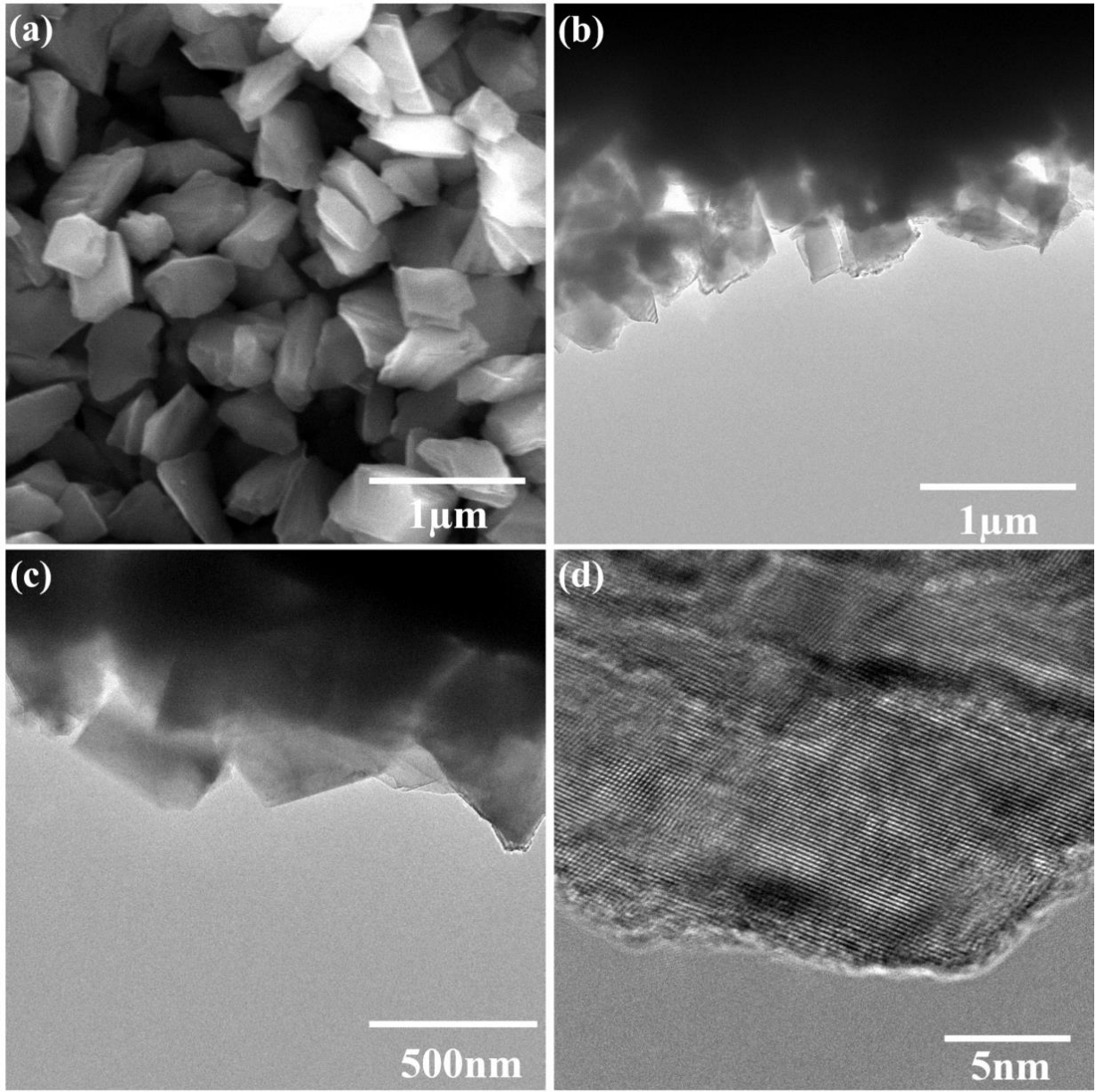

Figure 8. Micro/nano structure of diamond powder. (a) SEM image shows the microstructure of diamond powder. (b) TEM, (c) higher magnification TEM, and (d) HRTEM images of diamond nanosheets.

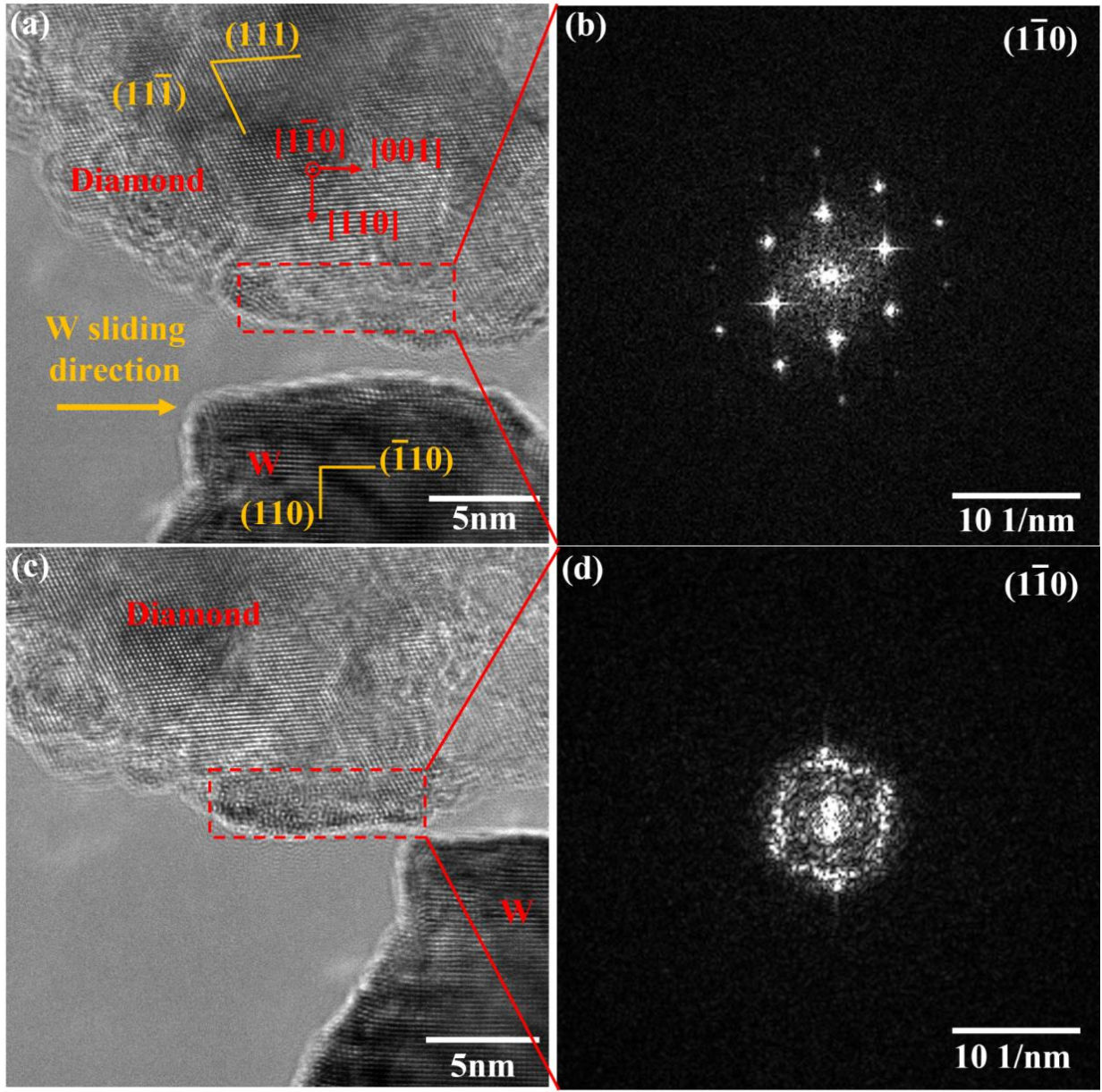

Figure 9. in-situ TEM observation of W/diamond friction. (a) HRTEM image showing diamond and tungsten crystalline structure before friction. (b) corresponding diffraction pattern of diamond in the selected area (the edge to be scraped with diamond) in (a). (c) HRTEM image showing diamond crystalline structure after friction. (d) corresponding diffraction pattern of diamond in the selected area (the edge scraped with diamond) in (c).

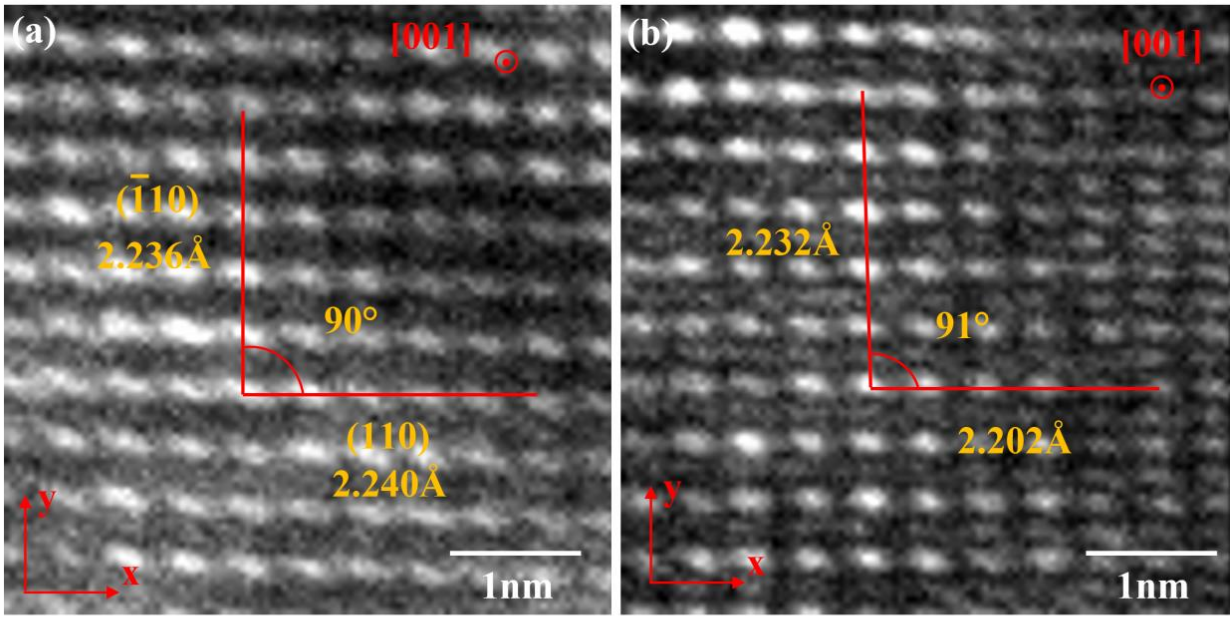

Figure 10. Lattice strain measurement of W caused by scractching with diamond. (a) HRTEM image showing W lattice structure before scraping with diamond. (b) HRTEM image showing W lattice structure when the interaction between diamond and W started.